# Evaluation of different methods for determining the angle of attack on wind turbine blades with CFD results under axial inflow conditions

H. Rahimi[a,b,], J.G Schepers[c,], W.Z Shen[d], N. Ramos García[d], M.S. Schneider[e], D. Micallef[f], C.J. Simao Ferreira[g], E. Jost[h], L. Klein[h], I. Herráez[i]

[a]*ForWind, Institute of Physics, University of Oldenburg, Oldenburg, Germany*
[b]*Fraunhofer Institute for Wind Energy Systems (IWES), Oldenburg, Germany*
[c]*ECN Wind Energy Technology, Petten, Netherlands*
[d]*Department of Wind Energy, Technical University of Denmark, Lyngby, Denmark*
[e]*German Aerospace Center (DLR), Institute of Aeroelasticity, Gttingen, Germany*
[f]*Department of Environmental Design, University of Malta, Msida, Malta*
[g]*Department of Wind Energy, Delft University of Technology, Delft, Netherlands*
[h]*Institute of Aerodynamics and Gasdynamics, University of Stuttgart, Germany*
[i]*Faculty of Technology, University of Applied Sciences Emden/Leer, Emden, Germany*

**Abstract**

This work presents an investigation on different methods for the calculation of the angle of attack and the underlying induced velocity on wind turbine blades using data obtained from three-dimensional Computational Fluid Dynamics (CFD). Several methods are examined and their advantages, as well as shortcomings, are presented. The investigations are performed for two 10MW reference wind turbines under axial inflow conditions, namely the turbines designed in the EU AVATAR and INNWIND.EU projects. The results show that the evaluated methods are in good agreement with each other at the mid-span, though some deviations are observed at the root and tip regions of the blades. This indicates that CFD results can be used for the calibration of induction modeling for Blade Element Momentum (BEM) tools. Moreover, using any of the proposed methods, it is possible to obtain airfoil characteristics for lift and drag coefficients as a function of the angle of attack.

**Nomenclature**

| | | |
|---|---|---|
| $\rho$ | Air density [kg m$^{-3}$] | |
| $\Gamma$ | Circulation [m$^2$ s$^{-1}$] | |
| $\phi$ | Inflow angle [°] | |
| $\Omega$ | Rotational speed [rad s$^{-1}$] | |
| $\theta$ | Twist angle [°] | |
| $a$ | Axial induction factor at each section [-] | |
| $a'$ | Tangential induction factor at each section [-] | |
| $C_l$ | Lift coefficient [-] | |
| $C_d$ | Drag coefficient [-] | |
| $U_\infty$ | Free stream wind speed at hub height [m s$^{-1}$] | |
| $U_{ind}$ | Induced velocity at each section, $U_{ind} = (U_\infty - U_{rotorDisk})$ [m s$^{-1}$] | |
| $y^+$ | Non-dimensional wall distance for a wall-bounded flow [-] | |

| | |
|---|---|
| *r* | Radial coordinate [m] |
| *R* | Rotor radius [m] |
| $V_{rel}$ | Relative velocity at each section [m s$^{-1}$] |
| *AoA* | Angle of Attack [°] |
| *BEM* | Blade Element Momentum |
| *CFD* | Computational Fluid Dynamics |
| *TSR* | Tip Speed Ratio |

**1. Introduction**

The traditional approach in wind turbine design codes used to simulate the aerodynamic behavior of wind turbines is the Blade Element Momentum (BEM) theory. The low computational cost of BEM, which is the result of many simplifications (e.g. steady and two dimensional flow), makes this an affordable approach even for the more than 7 million time steps which wind turbine manufacturers needs to perform in the calculation of the design load spectrum for certification [1]. In the BEM theory, the aerodynamic forces are interpolated from sectional airfoil characteristics such as $C_l$ and $C_d$ as a function of the Angle of Attack (AoA). Hence, the outcome of a BEM code depends heavily on these airfoil characteristics and AoA which are derived from two dimensional (2D) steady wind tunnel experiments, viscous-inviscid panel codes or Computational Fluid Dynamics (CFD) simulations [2].

In BEM, the assumption of 2D steady flow might be valid for the mid-span region and steady uniform inflow. Nonetheless, on a rotating blade, in a fluctuating 3D wind flow, which leads to a misalignment of the velocity vector with respect to the rotor plane and near the thick root section and the blade tip where the flow is highly complex and 3D, this assumption is not valid anymore. Therefore the complex 3D flows cannot be captured accurately by the 2D airfoil characteristics and basic BEM theory [3]. Hence empirical correction models are generally added to a BEM model to mitigate its lack of accuracy when specific phenomena are observed, such as: 3D effects [4, 5, 6, 7], dynamic stall [8, 9] or skewed wake [1, 10, 11].

However, due to the empirical nature of most engineering models, which are often based on findings obtained by rather small experimental wind turbines [12], their validity for the modeling of large rotors is currently under discussion [13, 1, 10, 11]. In the wind turbine research community, there is a continuous effort for improving engineering add-ons in BEM. An example is the EU project AVATAR [14], which aims at improving and validating aerodynamic models, and to ensure their applicability for designing 10MW+ turbines [13]. One could think of improving the accuracy of these correction models for the BEM codes from measurements using hot wires or Particle Image Velocimetry (PIV) [15, 16, 17, 18]. However, these measurements are very expensive in terms of both time and cost and there are often limited by spatial and temporal resolutions. An alternative method would be the usage of high fidelity numerical tools like CFD [19]. CFD is increasingly used to re-investigate the modeled phenomena for large rotor scales. Efforts made by Rahimi et al. [20, 19, 10, 21] showed that the high fidelity of CFD simulations can help to improve the skewed wake engineering model for large rotor size turbines. Another example is the work of Yang et al. [20] and Schneider et al. [21] related to stall delay models. Nevertheless, in all these works one of the major limitations for the improvement of these correction models using measurements or 3D CFD lies in the many uncertainties involved on the definition of AoA and underlying axial and tangential induced velocity for 3D flows.

The AoA is a 2D steady concept defined as the angle between the oncoming flow velocity and the airfoil chord, as shown in Figure 1. The AoA is not the angle measured with a probe/PIV or calculated with CFD ahead of the airfoil at a monitor point. In reality, trailing vortices due to variations in blade circulation alter the AoA. The local inflow velocity is influenced by the axial induction resulting from the momentum equilibrium. As such, if the influence of the presence of the airfoil, i.e. the influence of bound circulation is removed from any point of the flow field, the AoA is obtained, and thus the induced velocity.

Several methods have been introduced so far for calculating the AoA at the rotor in axial inflow conditions using 3D CFD simulations, such as inverse BEM [22], Average Azimuthal Technique (AAT) [3, 23] and the 'Shen' models from [24, 25]. However, there are very few studies such as Guntur et al [26] examining all of these methods for different conditions.

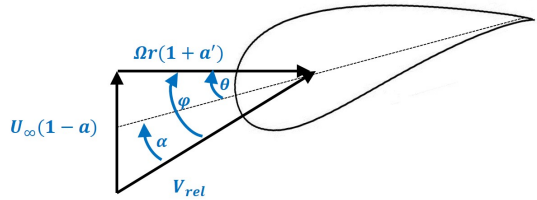

Figure 1: Schematic representation of the relative velocity $V_{rel}$, axial velocity $U_\infty(1-a)$, tangential velocity $\Omega r(1+a')$, AoA ($\alpha$), inflow angle ($\phi$) and twist angle ($\theta$).

The goal of the present work is therefore to assess available methods to extract AoA from CFD simulations, so that CFD results can be used for calibration of induction modelling in BEM. For the first step the axial inflow conditions without wind shear is considered.

For the numerical simulation, the EllipSys3D CFD solver [27, 28, 29] is used. The EllipSys3D code is a multi-block finite volume discretization of the incompressible Reynolds Averaged Navier-Stokes (RANS) equations in general curvilinear coordinates [29]. For this purpose, numerical simulation of two 3-bladed wind turbines, namely, the 10MW AVATAR [14] and the INNWIND.EU/DTU 10MW [30] turbines, by means of geometrically resolved CFD under uniform and axial inflow conditions are used. The free-stream velocities considered in this study are 6 and 9 $ms^{-1}$ with the rotor speeds of 6 and 7.3 RPM, respectively. In order to reduce the aerodynamic uncertainties, and to have the most clean comparison between the different models, CFD computations are performed for an isolated rotor only (neglecting the effect of tower and nacelle) in a uniform axial flow (no wind shear), excluding all aeroelastic effects. The operational conditions set for the turbines are the operating controlled conditions described in their reference manual which are publicly available at their website [14, 30]. In Table 1 the main parameters of the considered turbine are listed.

In the simulations, the pressure/velocity coupling is enforced through the Semi-Implicit Method for Pressure-Linked Equations (SIMPLE) algorithm [31]. The scheme which is used for this work is, QUICK scheme for the differencing scheme of the convective terms in the momentum equations, and first order upwind for the turbulence equations.

The computations are performed using the k-$\omega$ SST turbulence model [32] and the Drela/Giles version of the $E^n$ transition model [33]. All the computations are performed in a steady state mode using a rotating reference frame attached to the rotor blades. This means that the grid is not rotating physically and the Navier-Stokes equations are expressed in the polar rotating reference frame.

The computational grid has a spherical shape and is fully structured based on hexahedral cells. The grids are generated with DTU surface grid tools and the HypGrid3D code. The total grid size accounts to 14 million cells. It has 256 cells around the airfoils, 129 cells in the span-wise direction for each blade, and 129 cells in the wall normal direction. The $y^+$ values at the surface are kept below 1 everywhere on the blade surface with a cell size of $2\times10^{-6}$ normal to the wall.

It should be noted that the present work serves also as a verification of several methods since main participants used the same methods so that the correct implementation could be checked. Moreover, this contribution could be regarded as a survey on the assessment of each method under uniform and axi-symmetric inflow conditions.

## 2. Methods

Several techniques have been implemented for the calculation of the induction factor and AoA so far. In the following, these techniques are presented:

- Inverse BEM method: The computed or measured load distribution is used as an input to estimate the induction and AoA using the general BEM theory formulation [34, 35, 22]. This technique may be expected to give reasonable results in axial flow conditions. However due to the fact that BEM theory is one-dimensional, when it comes to flow separation or yawed conditions the accuracy of this model is questionable [1, 10, 11]. This also holds for the root and tip where the flow shows its 3D nature. In addition, just like in BEM, the simple momentum theory does not hold anymore when the axial induction factor becomes larger than approximately 0.4. Therefore, several empirical relations between the thrust coefficient and the induction factor have been proposed based on measurements in order to correct for this effect [36, 22] and it is essential to use such a correction in the inverse BEM, just like the BEM itself [21]. Hence the obtained AOA may be influenced by these factors.

| Turbine parameter | INNWIND.EU/DTU | AVATAR |
|---|---|---|
| Rated power (MW) | 10 | 10 |
| Rotor diameter (m) | 178.3 | 205.8 |
| Axial induction | 0.3 | 0.24 |
| Rotor speed (RPM) | 9.6 | 9.6 |
| Tip speed (ms$^{-1}$) | 90 | 103.4 |
| Hub height (m) | 119 | 132.7 |

Table 1: Basic reference turbine characteristics for the INNWIND.EU/DTU and AVATAR rotors.

Measurements or CFD results are supposed to be used as the validation materials for BEM codes and to improve the uncertainties in BEM. Therefore it would be better if the inverse BEM method would not be a part of the validation materials. However, it is also interesting to look at the validity of the general BEM theory formulation at different conditions.

- Average Azimuthal method (AAT): This technique is based on the annular average values of the axial velocity by using the data at several upstream and downstream locations [3, 23]. Figure 2 shows a schematic example of the annulus upstream monitoring points (in blue) and the annulus downstream monitoring points (in green) at a given radial location. Once the velocity field at each point for each annulus ring is known, it will be averaged over the annulus ring. Afterwards, the value of the velocity at the rotor plane is estimated by interpolating the upstream and downstream annulus ring averaged velocities. By knowing the induced velocity at the rotor plane, and the blade twist, the AoA for a given radial location can be found by:

$$\alpha = tan^{-1}\left(\frac{V_{averaged}}{r\Omega(1 + a')}\right) - \theta \qquad (1)$$

The shortcoming of this method lies in the fact that this model is only valid for axial conditions (or for the mean value of axial induction factor over one rotation) and it can not capture the dynamics of the induced velocity of yawed flows. Furthermore, since the method is based on averaged data, many input points have to be sampled, and therefore the computational cost might be high as compared to other methods. It is also noted that the results might depend on the positions of the monitor point and the interpolation algorithm. Finally, AAT provides an annular averaged induced velocity which is known to differ from the local induced velocity near the tip.

- 3-Point method: This method first introduced by Rahimi et al. in [10] uses three points along the chord length on each side of a particular section. This great simplicity is the main advantage of this method, which makes the calculation of the AoA ($\alpha$) very straightforward. In Figure 3, a schematic representation of this method is presented. By choosing three points at each section, the influence of bound circulation as well as the upwash and downwash effect is eliminated. In addition, unlike the AAT method, this method is able to reproduce the dynamic behavior of the induction and AoA (local induced velocity and AoA) for each azimuthal position and also near the tip and root of the blade which is very important for the yawed flow. This has been shown in previous works with a comparison to a lifting line free vortex wake code [10].

For this method, three points on each side of a particular section which is modeled as an airfoil will be assigned. These points are located in the chord-wise direction, at 25, 50 and 75 % of chord length along the airfoil at both sides. In the axial direction, they are located approximately 1.0×Chord away from the airfoil which means that their location are varying along the blade span and dependent on the local chord. The idea is that these

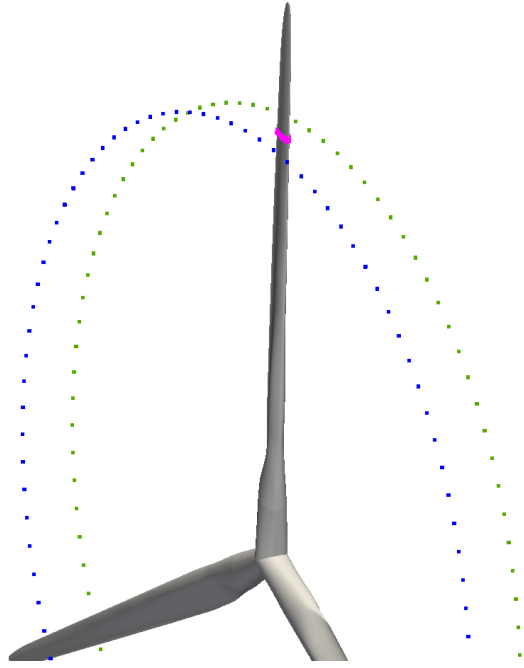

Figure 2: Average Azimuthal Technique (AAT) for the calculation of axial induction factor.

points cancel out the effect of the bound vortex and reduce the effect of flow separation on the resulting induced velocity. The procedure of extracting the induced velocity is as following:

Step 1: For each pair of points at upstream and downstream locations namely, (P1,P4), (P2,P5) and (P3, P6) the velocity is averaged independently by using an interpolation function. The three velocities $V_{1,4}$, $V_{2,5}$, $V_{3,6}$ are therefore obtained. Step 2: The estimated velocity which is induced at the blade section (red point in Figure 3) at the airfoil center can be approximated by simple averaging as: $V_{averaged} = (V_{1,4} + V_{2,5} + V_{3,6})/3$. Step 3: By knowing the induced velocity at the rotor plane, and the blade twist, the AoA for a given radial location can be found by:

$$\alpha = tan^{-1}\left(\frac{V_{averaged}}{r\Omega(1 + a')}\right) - \theta \qquad (2)$$

- Shen 1 method: The technique presented in [24] is a simple way to determine the AoA on a rotor blade. As input to this technique, the force distribution along the blade (typically projected along the chordwise and normal to the chordwise directions) and the velocity at a set of monitor points in the vicinity of the blade is assumed to be known from CFD computations or experiments. The AoA is determined by (1) estimating the lift force by projecting the force along the incoming and normal to the incoming directions; (2) calculating the bound vortex using the Kutta-Joukowski law; (3) calculating the induced velocity by the bound vortex using Biot-Savart's law; (4) computing the relative velocity at the monitor points by subtracting the induced velocity from the bound vortex; (5) computing the AoA from the relative velocity. This procedure continues until the convergence is reached. The position of the monitoring point is important for this method as the bound vortex is considered as a line vortex along the blade. The points cannot be chosen too close to the blade since this causes a singularity problem and the induced velocity approaches infinite. Therefore it is suggested that the monitoring point location be 2×chord away from the leading-edge in the rotational plane.

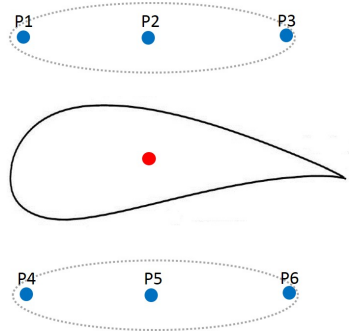

Figure 3: Schematic representation of the 3-Point method for the approximation of the velocities.

- Shen 2 method: In order to overcome the difficulty of singularity in the previous method, an alternative technique was presented in [25], where a distributed bound circulation along the airfoil/blade surface is used instead of the concentrated bound vortex at the force center. In this case, the monitor points can be chosen closer to the blade and it is shown that for this method the results are not depending on the location of monitoring points. It is suggested that the monitoring point location to be chosen from .5×chord away from the leading-edge in the rotational plane. Another advantage is that this method takes the chordwise variation of aerodynamic forces into account which is neglected when the vorticity is concentrated in a bound vortex. Additionally, this method is not iterative. However, the difficulty of using this method is to find the separation point (SP) where the local circulation changes sign [25].

- Ferreira-Micallef method: This method assumes that the flow around a blade section is 2D, incompressible and irrotational. The velocity at any point can be decomposed into three parts; (i) the velocity induced by free vorticity in the flow near the blade section, (ii) the induced velocity due to bound vorticity and (iii) the freestream wind velocity.

In this approach, the free vorticity and the vorticity bound to the surface of the airfoil are approximated by point vortices with a Rankine vortex distribution (see [37]), with a strength $\Gamma_\omega = \omega_z \Delta x \Delta y$. Figure 4 shows a schematic of the free vortex points distributed at each grid point. Bound vortex points of strength $\Gamma_{b_i}$ are placed on the blade surface, where $i$ is the index of the point vortex. The control points for the evaluation of the system of equations are collocated in the vicinity of the blade section, as represented in Figure 5.

The velocity induced at a point $(x, y)$ by a point vortex located at $(x_\omega, y_\omega)$ are given by Equations 3 and 4, where $u_\omega$ is the velocity induced in x-direction and $v_\omega$ is the velocity induced in y-direction.

$$u_\omega = \frac{\Gamma}{2\pi} \frac{y - y_\omega}{(x - x_\omega)^2 + (y - y_\omega)^2} \tag{3}$$

$$v_\omega = -\frac{\Gamma}{2\pi} \frac{x - x_\omega}{(x - x_\omega)^2 + (y - y_\omega)^2} \tag{4}$$

For simplification, we will define $(u_b, v_b)$ as the velocity induced by a point vortex bound to the surface of the blade section and $u_\omega, v_\omega$ as the velocity induced by a free point vortex. We can also define that the velocity induced by a surface bound vortex $i$ at a control point $j$ can be defined as $\left(u_{b_{ji}} = a_{ji}\Gamma_{bi}, v_{b_{ji}} = d_{ji}\Gamma_{bi}\right)$, where $a_{ij}$ and $d_{ij}$ are determined from Equation 3. For each control point $j$, the velocity $(u_j, v_j)$ is defined according to Equations 5 and 6, where $\sum u_\omega$ and $\sum v_\omega$ are the components of velocity induced by all free vortices and $\sum u_{b_{ji}}$ and $\sum v_{b_{ji}}$ are the components of velocity induced by all surface bound vortices. $(U_l, V_l)$ are the components of the local wind velocity at the blade's quarter chord.

$$u_j = \sum u_\omega + U_l + \sum u_{b_{ji}} \tag{5}$$

$$v_j = \sum v_\omega + V_l + \sum v_{b_{ji}} \tag{6}$$

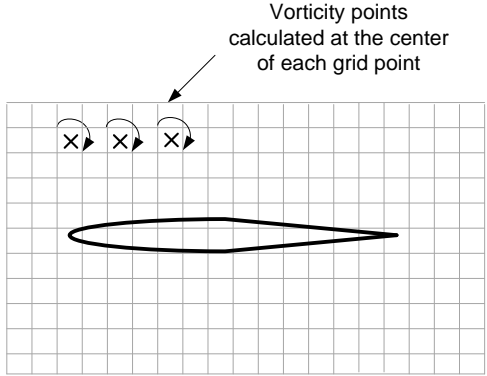
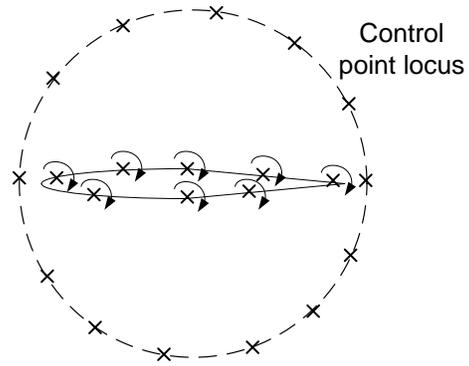

Figure 4: Schematic of free vortex points distributed at each grid point.

Figure 5: Schematic of the point vortices bound to the blade's surface and locus of the control points for velocity evaluation.

An over-constrained system can be solved using a least square method (this is done by means of Matlab's optimization toolbox). The velocity field from the full CFD simulation is used as input. This enables the determination of the local velocities $U_l$ and $V_l$ which can be used to find the angle of attack. The relative inflow angle is thus given by:

$$\phi = tan^{-1}\left(\frac{V_l}{U_l}\right) \quad (7)$$

Since the influence of tip vortices is 3D, some correction is needed especially near the root and tip.

- The Line average (LineAve) method determines the AoA by averaging the flow velocities along a symmetric, closed line around the rotor blade [38]. In the present study, a circle was chosen for this purpose as illustrated in Figure 6. The circle center is placed at the quarter chord position, where like in a lifting line representation the bound vortex is located. The idea is that the induced velocity at opposed points on the circle extinguish each other. By averaging the flow velocities along the circle, the influence of bound circulation is eliminated and the local inflow velocity and AoA can be determined.

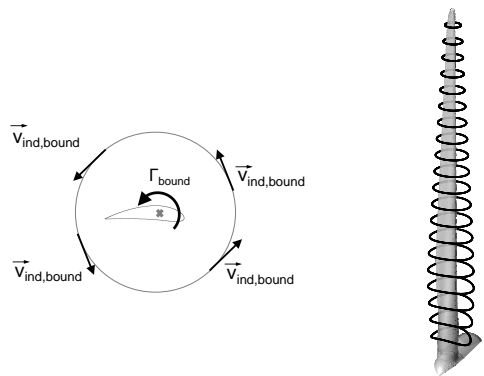

Figure 6: Line average method - 2D and 3D [38].

Since it needs to be solely ensured that the shapes are symmetric to the quarter chord point, in theory a variety of closed shapes are possible. However, in earlier works [38] the circle shape provided convincing results. The local circle radius is varied along the blade span and chosen dependent on the local chord length $c$. In this study, $r_{circ} = 1c$ is used. A verification and assessment of the method is performed in [38].

- Herráez Method: This method obtains the undisturbed flow rotor velocity by extracting them directly from a position in the rotor plane where the influence of the blade bound circulation from each blade is canceled out by the other blades [39]. In the case of axi-symmetric, homogeneous inflow, this position corresponds to the bisectrix of the angle between two arbitrary blades. For a wind turbine with 3-blades, the undisturbed velocity can be obtained along the radial traverses located 60° ahead and behind an arbitrary blade, as shown in Figure 7. In the case of a 2-bladed rotor, the velocities should be probed at the radial traverses located 90° ahead and behind an arbitrary blade. The difference between the free stream velocity and the axial velocity component obtained in this way for each radial position corresponds to the local axial wake induction. The local tangential velocity extracted from the radial traverse corresponds to the local tangential wake induction after changing its sign. It is demonstrated in Ref [39] that this method can predict the AoA satisfactorily from the root until at least 92% of the blade length. For larger spanwise positions, other methods, like e.g. Shen are preferred. The main advantage of this method is its simplicity, which makes the calculation of the AoA very straightforward. Furthermore, in opposition to other methods (see section 3.1), it is not sensitive to input parameters like engineering correction models, monitoring point location, etc. The main limitation of the method is that its use for non-axi-symmetric or inhomogeneous inflow becomes much more complicated because of the dependence of the blade bound circulation on the azimuthal blade position.

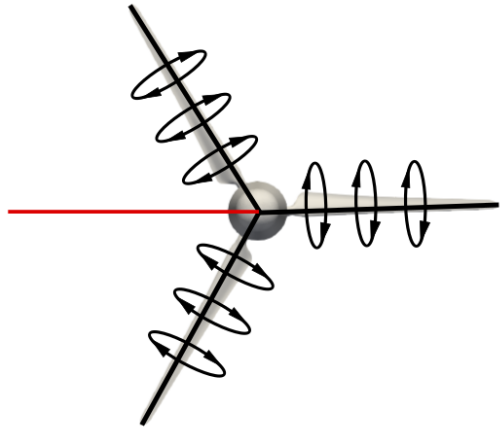

Figure 7: Three bladed rotor with axial and homogeneous inflow. The bound circulation distribution is the same for all blades. The red line represents a position where the influence of the bound circulation from all blades is cancelled out and the undisturbed velocity field can be probed.

## 3. Results

In this section, the results from the above mentioned methods for the calculation of the AoA and underlying induced velocity are presented for the INNWIND.EU/DTU and AVATAR turbines at two wind velocities of 6 $ms^{-1}$ and 9 $ms^{-1}$. The selected cases endeavour to be representative of the results of both the turbines which have been investigated.

### 3.1. Sensitivity study on input parameters: a, AoA

As explained in Section 2, different methods could have dependency on various input parameters e.g: the monitoring point location, turbulent wake correction or interpolation method etc. In this section, a sensitivity study on the input parameters for the AAT, Shen 1 and Shen 2 methods is presented. 3P and Line Average method also show similar behavior as the AAT method, therefore, they are not presented here. For the Ferreira-Micallef and Herráez methods, no dependency from input-parameters is reported.

*3.1.1. AAT method:*

The AAT method is based on probing the velocity at the specific radial positions in planes (upstream and downstream) parallel to the rotor. Therefore, the averaged axial velocity can be affected by the distance of the probing points (annular plane) to the rotor. The original AAT method assumes that the probing elements are located on a annular plane with a fixed distance to the rotor plane for all radial positions. In this study, the probing elements at each radial positions are in an annular ring which is parallel to the rotor with a distance which is a factor of chord length. The chord dependent monitoring point location is believed to be a more meaningful choice since the chord strongly varies from root towards the tip. In Figure 8, the dependency of $U_{ind}$ and AoA on the monitor point location is presented. The monitoring point location is changed from 0.5 × chord to 4 × chord for different cases. From Figure 8 it can be seen that at the very tip ($r/R$ >0.85 ) and root (r/R <0.35 ) $U_{ind}$ is slightly dependent on the position of the monitoring point due to the fact that when the monitoring points are located to far away from the blade in the axial direction, they are not on the same streamline anymore (extension of the stream tube. Nevertheless the extracted AoA generally remains in a very close agreement.

Moreover, the dependency of *a* and the AoA on the azimuthal resolution of the AAT method for the INNWIND.EU-DTU turbine at 9 $ms^{-1}$ is presented in Figure 9. It can be seen that when more than 80 points are used, *a* and AoA remain in a close agreement. Hence, it can be concluded that the azimuthal resolution has little effect on the results. The inner plot in Figure 8b represents AoA at the outer span location in a smaller range. It should be mentioned that the points are distributed in equidistant spacing in annulus upstream and downstream of the rotor plane. However concentrating the points near the blade did not present any differences.

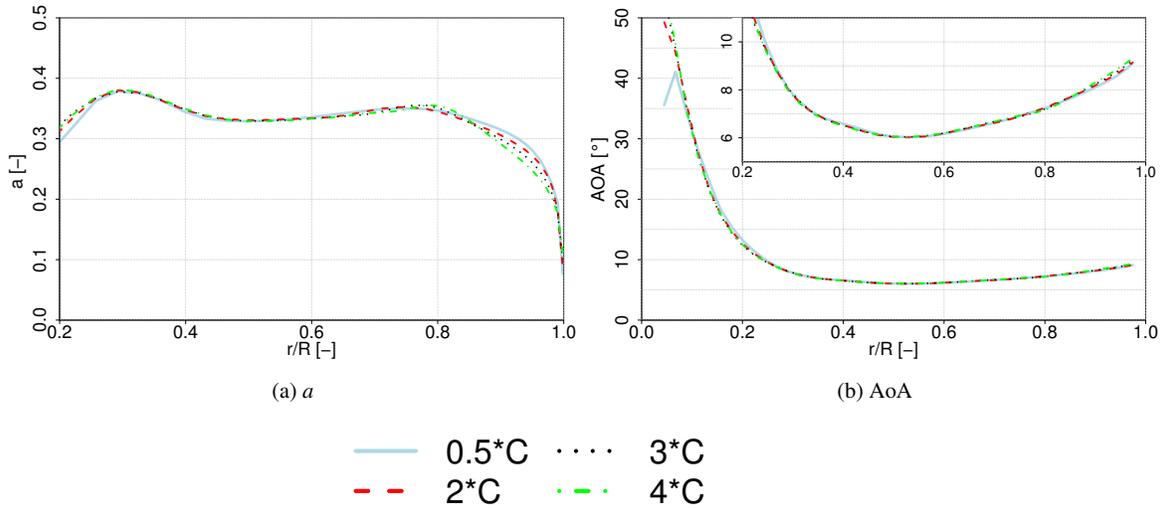

Figure 8: Dependency of *a* and AoA on the monitoring point location in the AAT method for the INNWIND.EU/DTU turbine at 9 $ms^{-1}$.

*3.1.2. Shen methods:*

The influence of changing the position of the monitoring points for Shen 1 and Shen 2 is studied in this section and presented in Figures 10 and 11, respectively. The monitoring point distance to the blade leading-edge is changed from 1 to 4 chord lengths. The results indicate that the dependency on the monitoring point locations has limited effect from $r/R$ =0.35 of rotor span onwards. However, a substantial dependency is observed in both *a* and AoA at the most inner-part of the blade as the monitoring points near the root are too close to the own blade or other blades. In addition, the dependency on the monitoring points near the root are higher for Shen 1 in comparison with the Shen 2 method. This is attributed to the fact that the method Shen 1 monitors points located close to the blade that are influenced by the singular bound vortex.

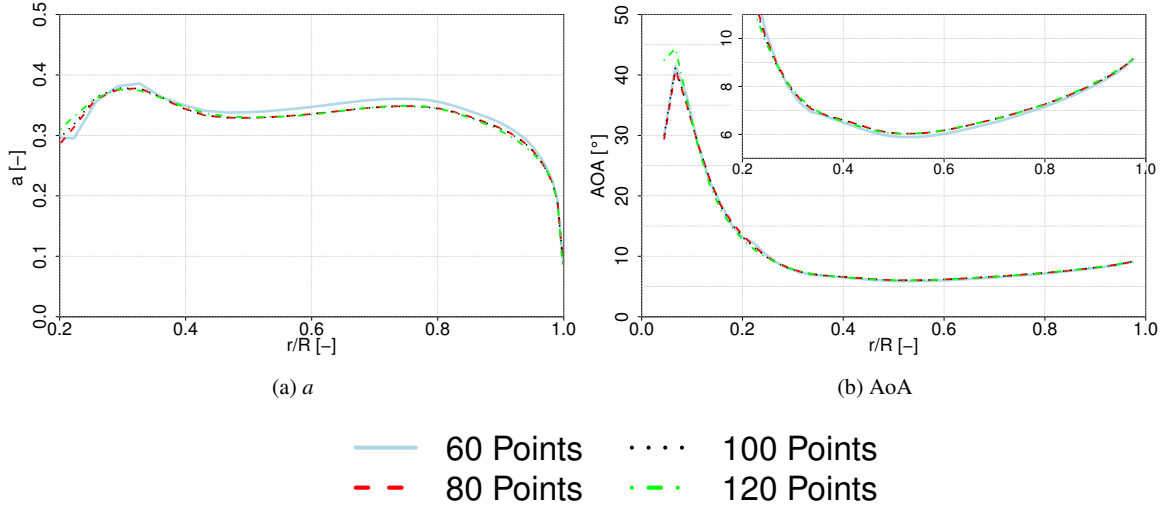

Figure 9: Dependency of *a* and AoA on the azimuthal resolution in the AAT method for the INNWIND.EU/DTU turbine at 9 $ms^{-1}$.

*3.1.3. Inverse BEM methods:*

In the inverse BEM method, the blade loads and the momentum equations of the BEM method were used to derive the AoA. There are several implementations of this method available such as [34, 35, 22], which for axial inflow conditions mainly differ on the correction method of high induction and the tip and root corrections. In the momentum equations, when the axial induction factor becomes larger than approximately 0.4 (denoted as $a_{critical}$), the simple momentum theory breaks down. Hence, several empirical relations between the thrust coefficient and *a* have been made using measurements to correct this effect [36, 22]. In Figure 12, the result of two implementations for $a_{critical}$ is presented. As it can be seen in Figure 12, although $a_{critical}$ has a certain impact on *a*, the AoA remains in a close agreement due to the fact that the tangential velocity plays a major role for the derivation of the AoA. The dependency of the results on different tip and root correction models is not presented here but can be found in Ref [24].

*3.2. Comparison of all the methods: a, AoA, $C_l$ and $C_d$*

In this section, the methods which were explained in Section 2 are used to extract the AoA and also the normalized aerodynamic force coefficients ($C_l$ and $C_d$). The results presented in this section are for the INNWIND.EU/DTU and AVATAR wind turbines at the inflow velocity of 9 $ms^{-1}$. From every method one result is presented. Thereto, the results from the sensitivity study in Section 3.1 have been analyzed and it is concluded that representative results are obtained when the monitor point is located at 4 times the chord length in front of the blade for the AAT, 3P and the Shen methods. For the AAT method 90 points in the annular ring are selected and for the inverse BEM the $a_{critical}$ is 0.33.

The numerical results are also compared with BEM calculations using FAST V8 [40], developed by the National Renewable Energy Laboratory (NREL). For the BEM calculation, polars from CFD computations are used. A uniform axial inflow without shear is used and no 3D correction model is applied in order to isolate the effects of the different airfoil polars. Figures 13a and 14a, the axial induction factor extracted from the CFD calculations using different methods is presented and also compared with the one from BEM. From the axial induction the AoA is extracted and presented in Figures 13b and 14b. Finally in Figures 13c, 14d, 13d and 14d the aerodynamic lift and drag coefficients are presented. From Figures 13 and 14 the following observation can be reported:

In general, in terms of axial induction factor (Figures 13a and 14a) a very good agreement is observed between all the methods at the mid-span (0.30 <*r/R* <0.85). However at the root (*r/R* <0.30) and at the tip of the blade (*r/R* >0.85) discrepancies can be observed. The reason for these discrepancies can be as following:

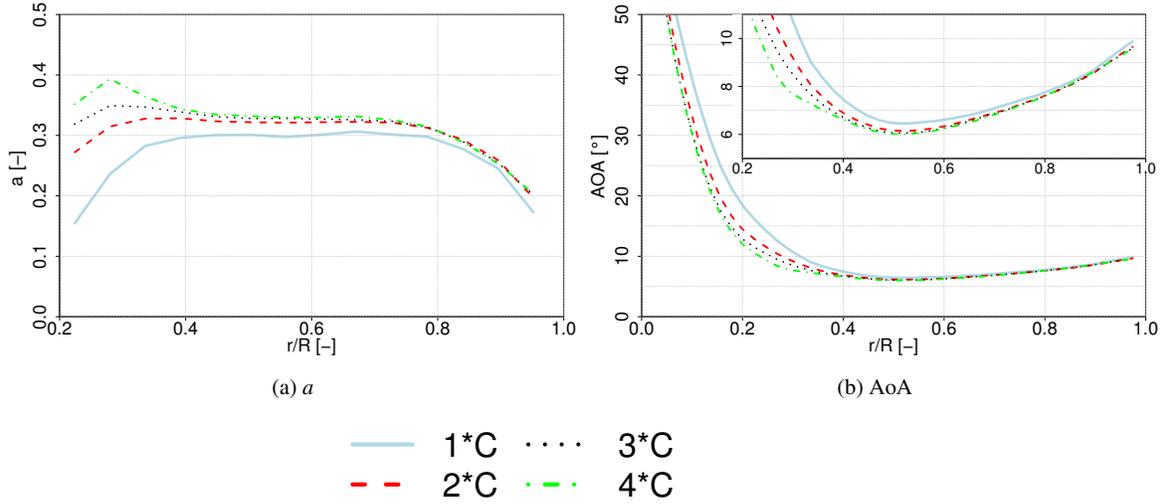

Figure 10: Dependency of *a* and AoA on the monitoring point location in the Shen 1 method for the INNWIND.EU/DTU turbine at 9 $ms^{-1}$.

1. Given that the methods employed consider various 2D sections along the blade, only the spanwise oriented bound circulation is assumed. In regions were 3D effects are predominant (such as root and tip), the bound circulation over the blade has some chordwise component which influences the blade induced velocity as well. Whereas the full CFD computations are able to capture the 3D flow as well as the loads in these zones along the blade, the methods under investigation here consider each section in a 2D sense. This is thought to be the main underlying reason for the discrepancies obtained in both root and tip. 3D corrections in these regions are necessary to account for the 3D character of bound circulation. The CFD computations provide a good idea of how blade bound circulation behaves at both tip and root. Various work has been conducted by [41, 42, 43] on this topic and this could be the basis for the envisaged corrections which are beyond the scope of this work.

2. The largest deviations are observed in the root area where flow is separated. As it can be seen in Figure 15, near the root area, the flow is largely separated, and a strong radial flow due to the Coriolis and centrifugal forces can be observed. In this area the positions of monitoring points play a major role one the calculated average value of induced velocity. When moving towards the mid-span however, the flow is less separated; therefore, the resulting monitored velocity is less sensitive to the probing location. In this sense, the methods using monitoring points far away from the blade surface are probably more accurate. Furthermore, methods relying on the Kutta-Joukowski law are not expected to be so accurate under such conditions.

3. For some methods, such as the line average and Ferreira-Micallef method, the monitoring points near the blade tip, are located closer to the tip vortex itself, therefore they predict different induced velocity as compared to other methods which have their monitoring points outside of the tip vortex.

4. When the monitoring points are located too far away from the blade in the axial direction, upstream and downstream monitoring points are not on the same streamline anymore (extension of the stream tube) therefore this could also cause some discrepancies in the tip.

5. At the tip, the AAT method gives an azimuthally averaged induction, the method of Herraez gives an value for the induction in the rotor plane, while all other methods represent the local induction.

In terms of AoA, first it is important to note that the AoA ($\alpha$) is the difference between inflow angle ($\phi$) and twist angle ($\theta$). The inflow angle as indicated in Figure 1 is the arctangent of the tangential velocity and the axial velocity. Figures 13b and 14b shows the AoA over the blade span. As the case of induction factor the level of agreement between the methods remains close at the mid span. However unlike the induction, the deviation between the methods is only observed at the root part ($r/R$ <0.30). Due to the high value of the rotational velocity in particular near the tip, differences in axial induction factor are "hidden" and so the AoA from the different methods will remain in a very close agreement all over the blade span.

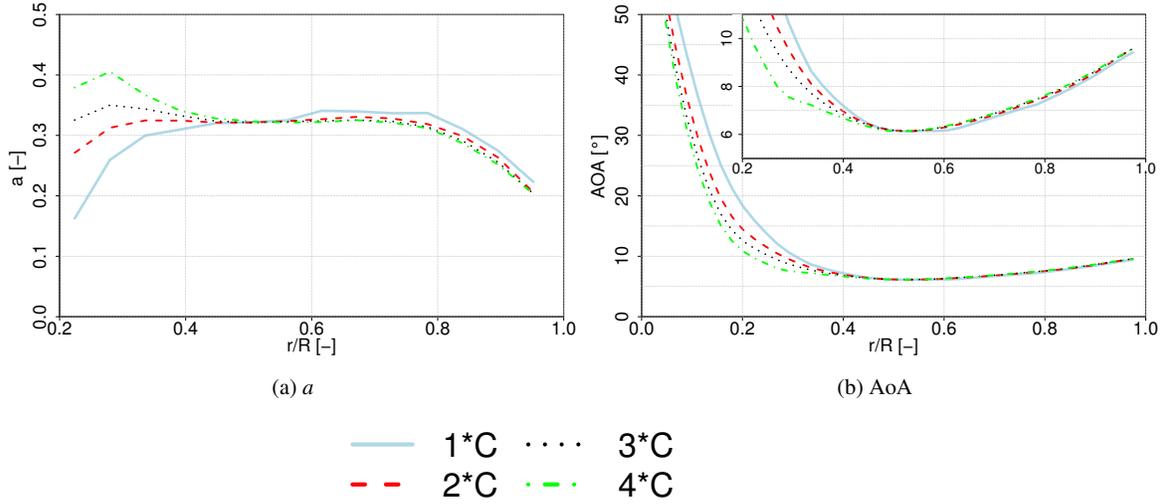

Figure 11: Dependency of $a$ and AoA on the monitoring point location in the Shen 2 method for the INNWIND.EU/DTU turbine at 9 $ms^{-1}$.

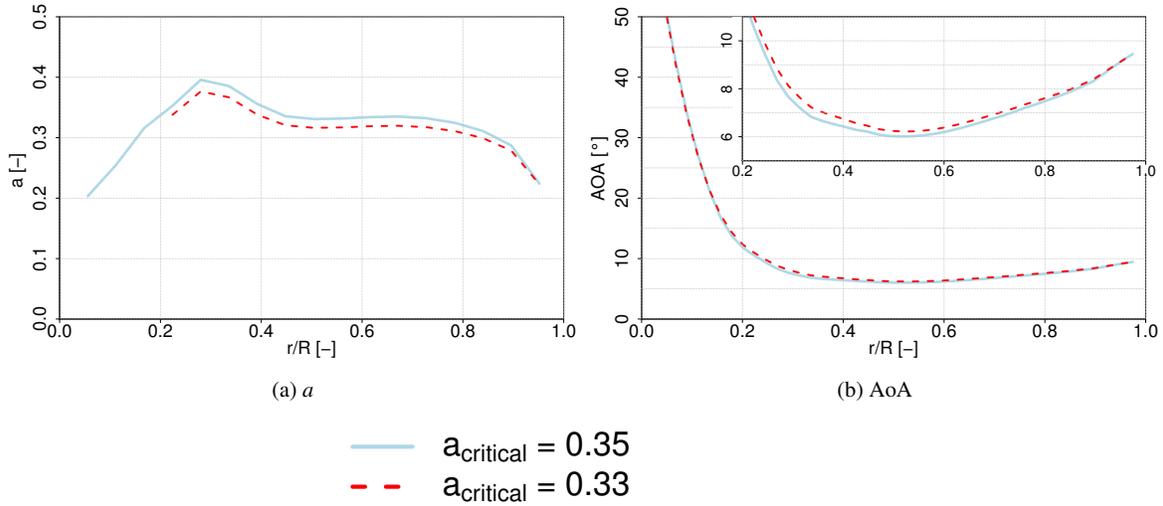

Figure 12: Dependency of $a$ and AoA on different corrections for high values of induction factor for the INNWIND.EU/DTU turbine at 9 $ms^{-1}$.

Finally in Figures 13c, 14d, 13c and 14d) the aerodynamic lift $C_l$ and drag coefficients $C_d$ are presented. For the mid-span the level of agreement between the methods remains close. At the very root and tip of the blade, $C_d$ shows some discrepancies. The thicker airfoil results in lower lift and higher drag with respect to the other parts of the blade. The $C_l$ is not affected by changes in the AoA. The rapid changes in $C_l$ and $C_d$ can be due to the root vortex or spanwise variations in the laminar to turbulent transition location. Also due to the Coriolis and centrifugal forces the extracted $C_l$ and $C_d$ are much higher than those extracted from 2D calculations without stall delay models.

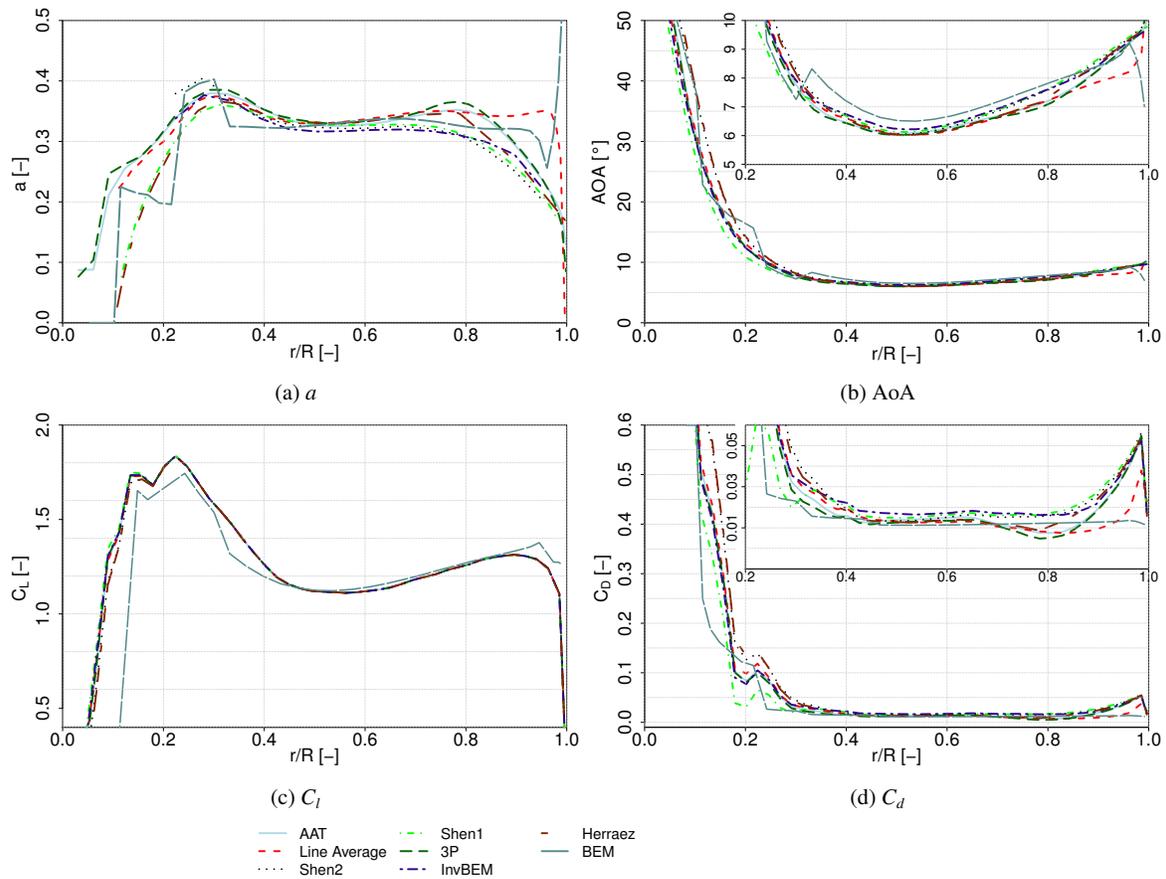

Figure 13: Comparison of all the considered methods in terms of $a$, AoA, $C_l$ and $C_d$ for the INNWIND.EU/DTU turbine at 9 $ms^{-1}$. For a better visibility the inner part of sub figures b,c and d represents the AoA, $C_l$ and $C_d$ at the outer span location in a smaller range.

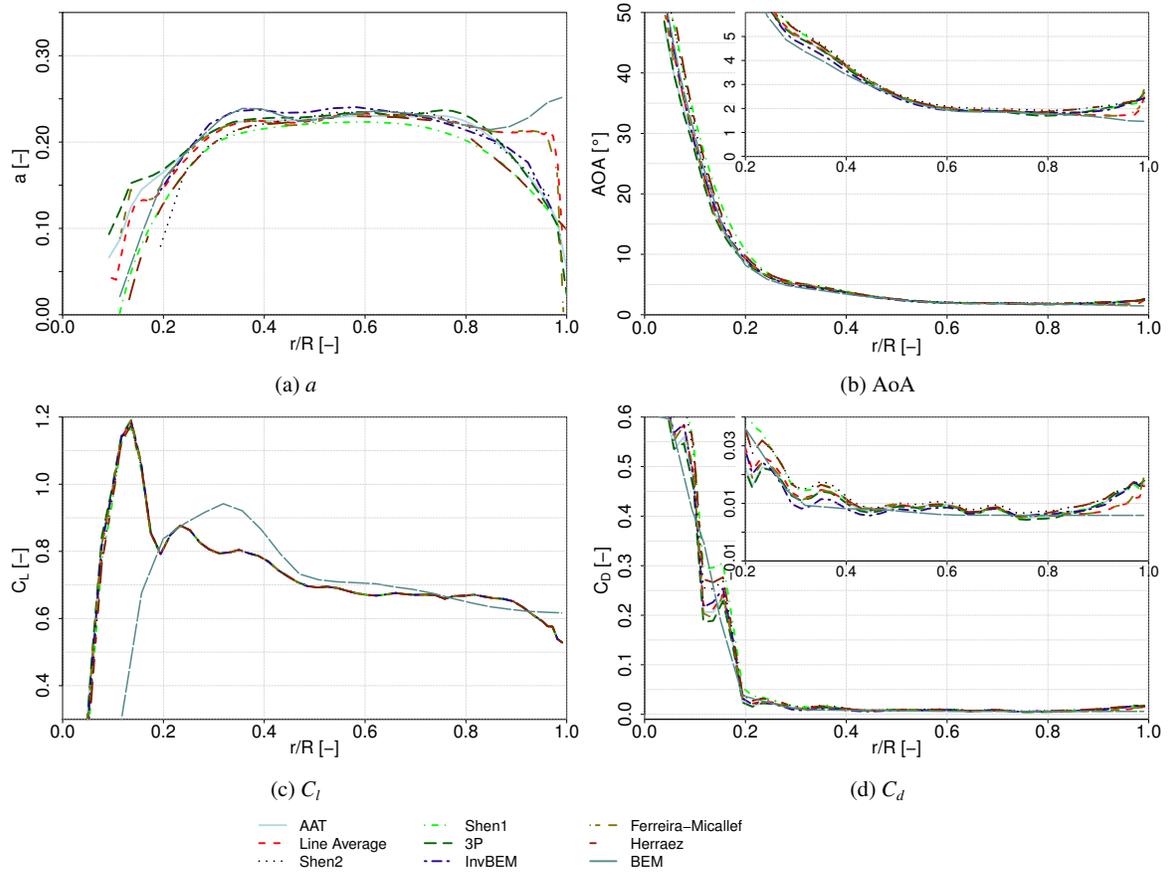

Figure 14: Comparison of all the considered methods in terms of $a$, AoA, $C_l$ and $C_d$ for the AVATAR turbine at 9 $ms^{-1}$. For a better visibility the inner part of sub figures b,c and d represents the AoA, $C_l$ and $C_d$ at the outer span location in a smaller y-range.

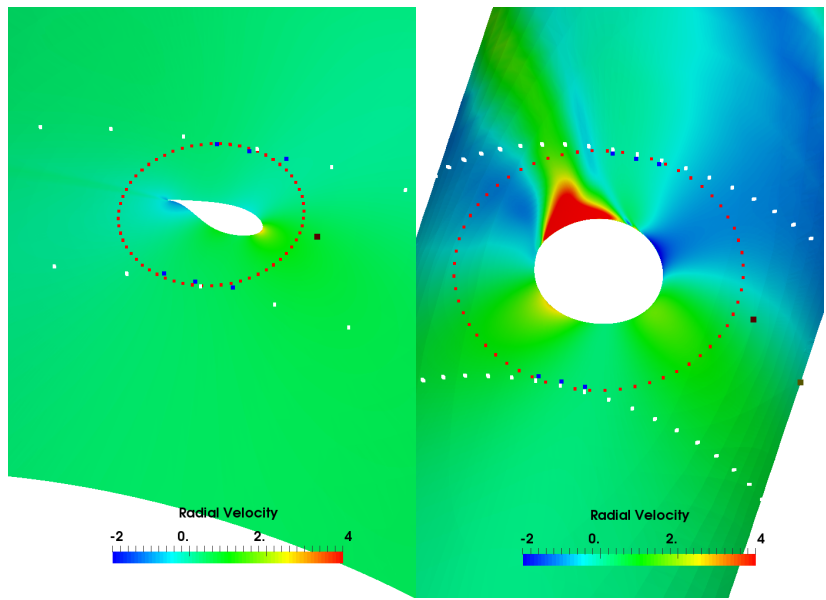

Figure 15: Schematic positions of the monitoring points for different methods at r/R=40 % (left) and r/R=10 % (right). AAT in white, 3-Point in blue, LineAve in red, Shen 1, 2 in brown and Herraez method in olive color.

## 4. Conclusion and future work

In this work, several methods for extracting the induced velocity and AoA from CFD computations in axial inflow conditions for flows past two 10MW class wind turbines, namely the EU AVATAR and INNWIND.EU/DTU turbines, are presented. The main outcomes of the work are as follows:

1. Most methods presented in this work have been used at least twice by different partners and give consistent results at mid-span. This finding is important new knowledge.

2. At the root and tip, 3D effects are predominant, the methods fail to account for velocity induced by chord-wise bound circulation. Therefore strong discrepancies have been observed in these regions showing a lack of reliability.

3. Whereas the methods make use of high fidelity data (such as that produced by CFD) as input, the underlying methodology in methods such as the LineAverage, Ferreira-Micallef, Shen 1, AAT and 3-Point relies on 2D theory.

4. When the monitoring points are located too far away from the blade in the axial direction, the up and downstream points are not on the same streamline anymore (extension of the stream tube). This causes discrepancies for the methods which are using the monitoring points.

5. In the region with the 3D flow, the location of the monitoring points leads to strong discrepancies on the resulting induced velocity.

6. Deviations near the tip have less impact on the resulting AoA due to the high value of the rotational velocity in particular.

7. Refinements to the methods need to come from a clear understanding of the behaviour of bound circulation in 3D flow dominated regions.

In spite of the fact that, even with experimental data, no direct validation is possible, the different methods showed a good agreement at the mid-span. This indicates that CFD results can be used for the calibration of induction modeling for BEM tools. For the outer part alternative strategies might be pursued. In future work, these methods will be tested in asymmetric inflow conditions which might enable the improvement of BEM induction modelling at e.g shear, yaw or turbulence inflow.

**Acknowledgments**

The authors would like to thank Niels N. Sørensen from DTU for providing the CFD simulations. The authors would also like to thank Elia Daniele from Fraunhofer IWES for his helpful discussions. This project has received partly funding from the European Union 7th framework program for research, technological development and demonstration under the grant agreement No FP7-ENERGY-2013-1/no 608396 (AVATAR project). This work has also partly supported by the Energy Technology Development and Demonstration Program (EUDP-2014, J. nr. 64014-0543) under the Danish Energy Agency.